\documentclass[prl,twocolumn,aps,superscriptaddress]{revtex4-1}
\usepackage{graphicx}
\usepackage{amssymb,latexsym,amsthm,amsmath}    
\usepackage{color}
\usepackage{subfigure}
\usepackage{changes}
\colorlet{Changes@Color}{orange}
\begin{document}

\title{Synchronization centrality and explosive synchronization in complex networks}

\author{A. Navas}
\affiliation{Center for Biomedical Technology, Univ. Polit\'ecnica de Madrid, 28223 Pozuelo de Alarc\'on, Madrid, Spain}
\author{J. A. Villacorta-Atienza}
\affiliation{Department of Applied Mathematics, Facultad de Ciencias Matem\'aticas, Universidad Complutense, Madrid, Spain}
\author{I. Leyva}
\affiliation{Center for Biomedical Technology, Univ. Polit\'ecnica de Madrid, 28223 Pozuelo de Alarc\'on, Madrid, Spain}
\affiliation{Complex Systems Group $\&$ GISC, Univ. Rey Juan Carlos, 28933 M\'ostoles, Madrid, Spain}
\author{J. A. Almendral}
\affiliation{Center for Biomedical Technology, Univ. Polit\'ecnica de Madrid, 28223 Pozuelo de Alarc\'on, Madrid, Spain}
\affiliation{Complex Systems Group  $\&$ GISC, Univ. Rey Juan Carlos, 28933 M\'ostoles, Madrid, Spain}
\author{I. Sendi\~na-Nadal}
\affiliation{Center for Biomedical Technology, Univ. Polit\'ecnica de Madrid, 28223 Pozuelo de Alarc\'on, Madrid, Spain}
\affiliation{Complex Systems Group  $\&$ GISC, Univ. Rey Juan Carlos, 28933 M\'ostoles, Madrid, Spain}
\author{S. Boccaletti}
\affiliation{CNR-Institute of Complex Systems, Via Madonna del Piano, 10, 50019 Sesto Fiorentino, Florence, Italy}
\affiliation{Embassy of Italy in Israel, Trade Tower, 25 Hamered St., 68125 Tel Aviv, Israel}

\begin{abstract}
Synchronization of networked oscillators is known to depend fundamentally on the interplay between the dynamics of the graph's units and the microscopic arrangement of the network's structure. For non identical elements, the lack of quantitative tools has hampered so far a systematic study of the mechanisms behind such a collective behavior. We here propose an effective network whose topological properties reflect the interplay between the topology and dynamics of the original network. On that basis, we are able to introduce the "synchronization centrality", a measure which quantifies the role and importance of each network's node in the synchronization process. In particular, we use such a measure to assess the propensity of a graph to synchronize explosively, thus indicating a unified framework for most of the different models proposed so far for such an irreversible transition. Taking advantage of the predicting power of this measure, we furthermore discuss a strategy to induce the explosive behavior in a generic network, by acting only upon a small fraction of its nodes.

\end{abstract}

\maketitle


One of the most intriguing processes in complex networks' dynamics is synchronization: the spontaneous organization of the network's units into a collective dynamics. This phenomenon is known to be related to a delicate interplay between the topological attributes of the network and the main features of the dynamics of each graph's unit \cite{Boccaletti06, Dorogovtsev08, Arenas08}. The conditions for synchronization in complex networks have been addressed by means of different approaches. For identical units, one of the most successful tools is, for instance, the {\it Master Stability Function} \cite{Pecora98}, which rigorously shows how the spectral properties of the graph influence the stability of synchronization \cite{Boccaletti06}. However, the general case of non-identical units is far more complicated, and often needs a numerical approach, where the topology-dynamics relationship can only be investigated within specific scenarios \cite{scenarios1,scenarios2,scenarios3,scenarios4,scenarios5}.

Such a connection between structure and dynamics of a network is of particular importance in the case of the recently reported explosive synchronization (ES), an irreversible and discontinuous transition to the graph's synchronous state. Originally, ES was described in all-to-all coupled ensembles of Kuramoto oscillators \cite{Kuramoto} for a specific distribution of natural frequencies \cite{Pazo05}. Later on, various kinds of degree-frequency correlations were found to be able to induce ES in networks of periodic and chaotic oscillators \cite{Gardenes11, Leyva12,PLi13,Skardal14}, or neural networks \cite{Chen13}. In addition, other microscopic mechanisms were proposed, based on diverse coupling strategies \cite{Leyva13_g, Leyva13_w, Zhang14}, or by introducing adaptive dynamics in a fraction of the network's units \cite{Zhang15}.

In this Letter, we propose the use of an effective network whose structure indeed reflects the interplay between the topology and dynamics of the original system. On that basis, we introduce the {\it synchronization centrality} as a measure to quantify the role of each node in the synchronization process. This measure allows us to provide a general explanation of the mechanisms underlying ES and revisit the main scenarios where such a behavior was previously reported. Finally, we formulate a criterion to induce explosive transitions by acting only on a small fraction of the network's nodes.

We start by considering a network of $N$ phase oscillators, whose instantaneous phases evolve in time according to the Kuramoto model \cite{Kuramoto}:
\begin{equation}
\label{kura1}
\dot{\theta}_{i}=\omega_i+\frac{\sigma}{\langle k\rangle}\sum_{j=1}^N A_{ij}\sin(\theta_j-\theta_i)\;\;\;i=1,...,N,
\end{equation}
where $\theta_i$ is the phase of the $i$-th oscillator, $\omega_i$ its natural frequency (chosen from a generic, known, distribution $g(\omega)$), $\sigma$ the coupling strength, and $\langle .\rangle$ stands for the ensemble average. The topology of the network is encoded in the adjacency matrix ${\bf A}$ ($A_{ij}=1$ if node $i$ is linked to node $j$, and $A_{ij}=0$ otherwise). The degree of a node is  $k_i=\sum_{j}A_{ij}$.
The level of synchronization is measured by the order parameter $r=\frac{1}{N}\langle\vert\sum_{i=1}^Ne^{\theta_i}\vert\rangle_T$, where $\vert .\vert$ and $\langle .\rangle_T$ denote module and time average, respectively.  Along this Letter, the network size is fixed to $N=1,000$ and the natural frequencies $\omega_i$ are randomly drawn from a uniform distribution in the interval  $[-0.5,0.5]$, unless otherwise specified.

As the coupling strength $\sigma$ gradually increases, system (\ref{kura1}) experiences a transition from an incoherent ($r\simeq 0$) to a synchronized state ($r\simeq 1$), a process often referred to as {\it path to synchrony} (PTS) \cite{Gardenes06}. In heterogeneous networks, the PTS is mainly dominated by the most connected nodes (or hubs) which actually act as synchronization seeds, and progressively recruit the other network's nodes. At variance, in homogeneous networks, the PTS is characterized by the emergence of coherent clusters growing around multiple synchronization seeds. Along the PTS, the first nodes that locally synchronize generally correspond to pairs of connected oscillators whose natural frequencies are closer, whereas the globally synchronized state emerges around those with natural frequencies close to the synchronization frequency $\Omega_s$.

While traditionally attention has been focused on the natural dynamics of each node, we now discuss a different approach, where the links prevail over the nodes themselves, and we use the frequency detuning $\Delta\omega_{ij}\equiv\vert\omega_i-\omega_j\vert$ between each pair of nodes as the key dynamical feature for the determination of the PTS.  To formalize our idea, let us introduce a change of variables $r_ie^{i\Psi_i}=\frac{1}{\langle k\rangle}\sum_{j\in \Gamma_i}e^{i\theta_j}$, where $r_i(t)$ is a local order parameter and $\Gamma_i$ is the set of neighbors of the node $i$. Substituting into Eq.~(\ref{kura1}) we obtain
\begin{equation}
\label{kura2}
\dot{\theta}_{i}=\omega_i+\sigma r_i\sin \left(\Psi_i-\theta_i \right).
\end{equation}
It follows naturally that the velocity difference is  $\dot{\Phi}_{ij}=\dot{\theta}_{i}-\dot{\theta}_{j}=\omega_i-\omega_j+\sigma\left[ r_i\sin(\Psi_i-\theta_i)-r_j\sin(\Psi_j-\theta_j)  \right]$.
As synchronization implies $\dot{\Phi}_{ij}=0$, the set of links through which synchronization may take place must fulfill
\begin{equation}
\label{condition}
\Delta\omega_{ij}\leq\sigma \left( r_i+r_j\right),
\end{equation}
which in fact relates local synchronization to the frequency detuning associated to the links, rather than to the microscopic structure of connections
around the nodes. 

\begin{figure}[h]
 \centering \includegraphics[width=0.48\textwidth]{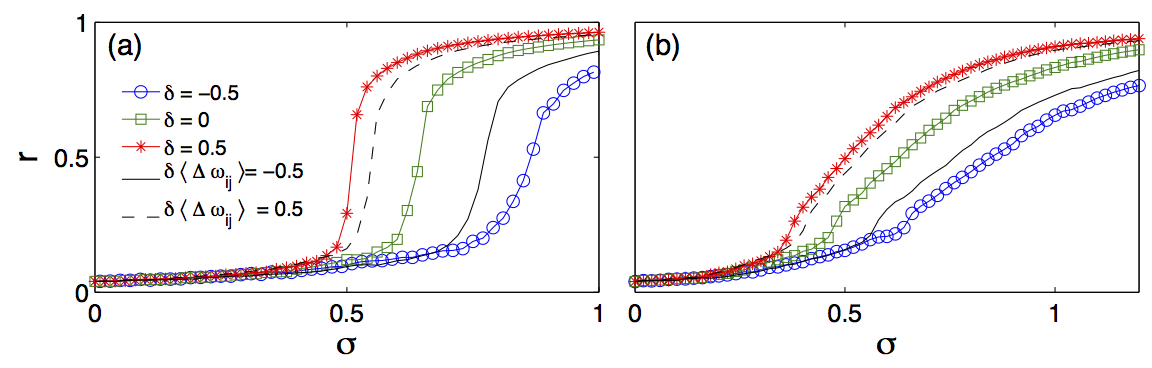}
  \caption{(Color online). Synchronization transition in perturbed (a) ER, $<k>=30$ and (b) SF $<k>=12$ networks, as compared with the original unperturbed system (black circles). Red triangles and blue squares shows that a positive (negative) perturbation of the adjacency matrix with $\delta=0.5$ ($\delta=-0.5$) reinforces (weakens) the local coupling of those links with large detuning, and results in an increase (frustration) of the level of synchronization.
  Furthermore, the local perturbation enhances/frustrates the synchronization more efficiently than a global perturbation over all links (dashed lines, where $\delta\Delta\omega_{ij}\rightarrow\delta\langle \Delta\omega\rangle$, being $\langle\Delta\omega\rangle$ the average over nonzero values of the detuning matrix).} \label{pert}
\end{figure}

To show further the role of frequency detuning, we consider the effect on synchronization caused by perturbing the adjacency matrix as $
 A_{ij}\rightarrow A_{ij}\left( 1+\delta\Delta\omega_{ij}\right)$.
Figure \ref{pert} reports the synchronization for (a) Erd\"os-Reyni (ER) $<k>=30$ and (b) scale-free (SF) networks generated by Barabasi-Albert algorithm with $<k>=12$. It can be seen an enhancement (frustration) of the synchronization as $\delta$ is increased (decreased).  Hence, positive (negative) values of $\delta$ potentiate (weaken) the strength of the couplings with $\Delta\omega_{ij}$,
forcing (preventing) pairs of nodes with large detuning to be effectively closer in natural frequencies. Notice that such a local perturbation of the adjacency matrix is more effective in promoting or delaying the PTS than a global perturbation equally acting on all links as shown by the corresponding dashed lines in Fig.~\ref{pert}.

\begin{figure}[t]
\centering \includegraphics[width=0.48\textwidth]{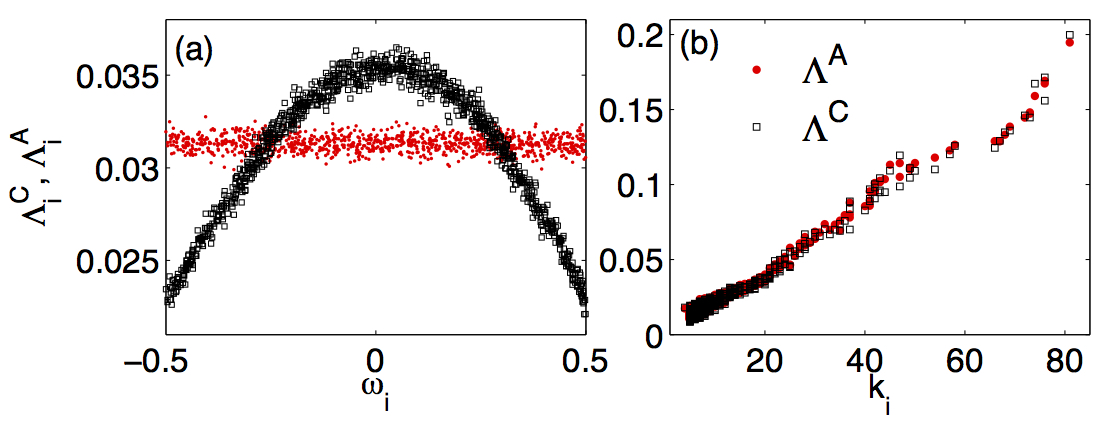}
\caption{(Color online) Comparison between synchronization centrality $\Lambda^C_i$ (black squares) and topological centrality $\Lambda^A_i$ (red dots). (a) ER networks, $\langle k \rangle$=50, $\Lambda^C_i$ and $\Lambda^A_i$ are reported {\it vs.} the nodes' natural frequencies $\omega_i$; (b) SF networks, $\langle k \rangle$=12, $\Lambda^C_i$ and $\Lambda^A_i$ are plotted {\it vs.} the node degrees $k_i$. All data refer to ensemble averages over 100 different network realizations.}
\label{LA_vs_LC}
\end{figure}

Motivated by Eq.~(\ref{condition}), we now merge the structural and dynamical information by introducing an \textit{effective network} characterized by the adjacency matrix ${\bf C}=\lbrace C_{ij} \rbrace$, where
\begin{equation}
C_{ij}\equiv A_{ij}\left( 1-\frac{\Delta \omega_{ij}}{\Delta\omega_{max}} \right),
\label{C}
\end{equation}
being $\Delta \omega_{max}$  the maximum possible detuning present in the system. In this way, the effective network exhibits the topology of the original network but enhancing those more synchronizable pairs of nodes according to  Eq.~(\ref{condition}).
We remark that, although we are here referring to the Kuramoto model of Eq. (\ref{kura1}), Eq.~(\ref{C}) can be applied much broadly to any
kind of oscillator's ensemble which can be associated to a set of well defined natural frequencies ${\omega_i}$.
Indeed, numerical results (not reported here) indicate that the main conclusions we will draw for Eq. (\ref{kura1}) are
in fact valid also for networks of chaotic oscillators, provided that the power spectrum of each unit is pronouncedly peaked around a unique
frequency, i.e. the units are strongly {\it phase coherent} \cite{boccalettiphysrepsynch}.

Now, in order to quantify the role of each node in the synchronization process,
we extract the most important nodes in the effective network defined by ${\bf C}$, i.e.
we calculate the eigenvector centrality \cite{Boccaletti06} of ${\bf C}$, and obtain the {\it synchronization centrality} vector $\bf\Lambda^C$.
The $i$-th component $\Lambda^C_i \ge 0$ provides a measure of the importance of the node $i$ in the effective network and quantifies its potential to behave as a seed of synchronization.

As a testing ground, Fig.~\ref{LA_vs_LC} shows the comparison between $\bf\Lambda^C$ and its topological counterpart $\bf\Lambda^A$, the eigenvector centrality extracted from the original adjacency matrix ${\bf A}$. For Erd\"os-R\'enyi (ER) networks \cite{Erdos}, the distribution of the components of the vector $\bf\Lambda^C$ as a function of the corresponding node's natural frequencies  shows the existence of many seeds of synchronization with natural frequencies close to $\Omega_s=0$  (see the black squares of Fig.~\ref{LA_vs_LC}(a)). This allows the  characterization of the connection between the microscale (detuning of the links) and the macroscale (emergence of global synchronization) of the system in a much better way than $\bf\Lambda^A$, whose components (red dots in Fig.~\ref{LA_vs_LC} a) are instead uniformly distributed. For heterogeneous scale-free (SF) networks \cite{Barabasi}, the synchronization seeds are the hubs and therefore $\bf\Lambda^C$ and  $\bf\Lambda^A$ provide essentially the same information, as it can be seen in Fig.~\ref{LA_vs_LC}(b).

\begin{figure}[t]
\centering  \includegraphics[width=0.43\textwidth]{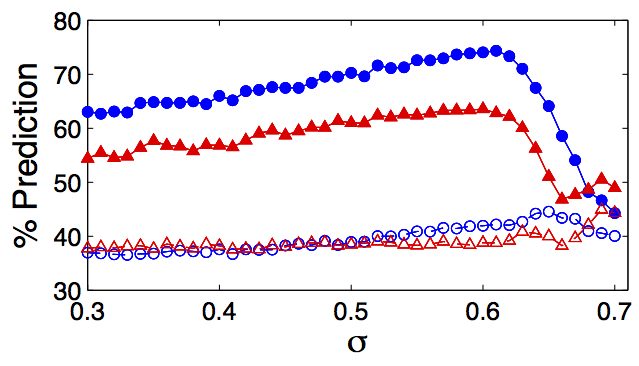}
\caption{(Color online). Average percentage of coincidence  between the third of the nodes with the highest (red) and lowest (blue) dynamical $\Lambda_i^S$ and synchronization $\Lambda_i^C$ centralities (solid symbols) or topological $\Lambda_i^A$ centralities (empty symbols). Calculations are performed on ER networks with same specification as in the Caption of Fig. \ref{LA_vs_LC}, and refer to 20 realizations of the network's topology and frequency distribution (see text for details).}
\label{pred}
\end{figure}

In order to further show how the effective network allows to predict the leading role of each node, we proceed with confronting our approach with a dynamical exploration of the system for the case of  ER networks. Precisely, we calculate the local synchronization matrix ${\bf S}=\lbrace S_{ij} \rbrace =\vert\langle e^{i\Delta\theta_{ij}}\rangle_t\vert$ (being $\Delta\theta_{ij}=\theta_{i}-\theta_{j}$) for a range of $\sigma$ values around the synchronization threshold. The eigenvector centrality of {\bf S}, denoted by $\bf\Lambda^S$, provides the actual dynamical centrality of each node in the synchronization process.
In Fig.~\ref{pred} we report the  percentage of coincidence  between the third of the nodes with the highest (lowest) dynamical and synchronization centralities ($\Lambda_i^S$ and $\Lambda_i^C$ respectively), and the percentage of coincidence between the dynamical and topological ($\Lambda_i^A$) centralities. It can be seen that the ranking based on
$\Lambda_i^C$ (curves with solid symbols) is able to predict up to 65$\%$ (75$\%$) of the nodes with the highest (lowest) synchronization centrality, while the topological centrality (curves with empty symbols) only detects at most the 30\%. The maximum of predictability is reached around the synchronization threshold and decreases rapidly due to the homogenization of the matrix {\bf S} for overcritical couplings.

As synchronization centrality reveals itself as a suitable measure to characterize the PTS, we move on
elucidating how this quantity helps us also to understand the microscopical mechanisms underlying ES. As it has been remarked in the introduction, ES can be induced when topology and dynamics are related in several specific ways  \cite{Pazo05,Gardenes11,Leyva12,Leyva13_g,Leyva13_w,Zhang13,PLi13,Skardal14,Zhang14,Zhang15}. Almost all these methods are based on a manipulation of the adjacency matrix and/or the links weights, such that in Eq.~(\ref{kura1}) A$_{ij}$ is replaced by a certain matrix $\Omega_{ij}$ which usually correlates the structural and dynamical features of the network.
\begin{figure}[t]
\centering
\includegraphics[width=0.5\textwidth]{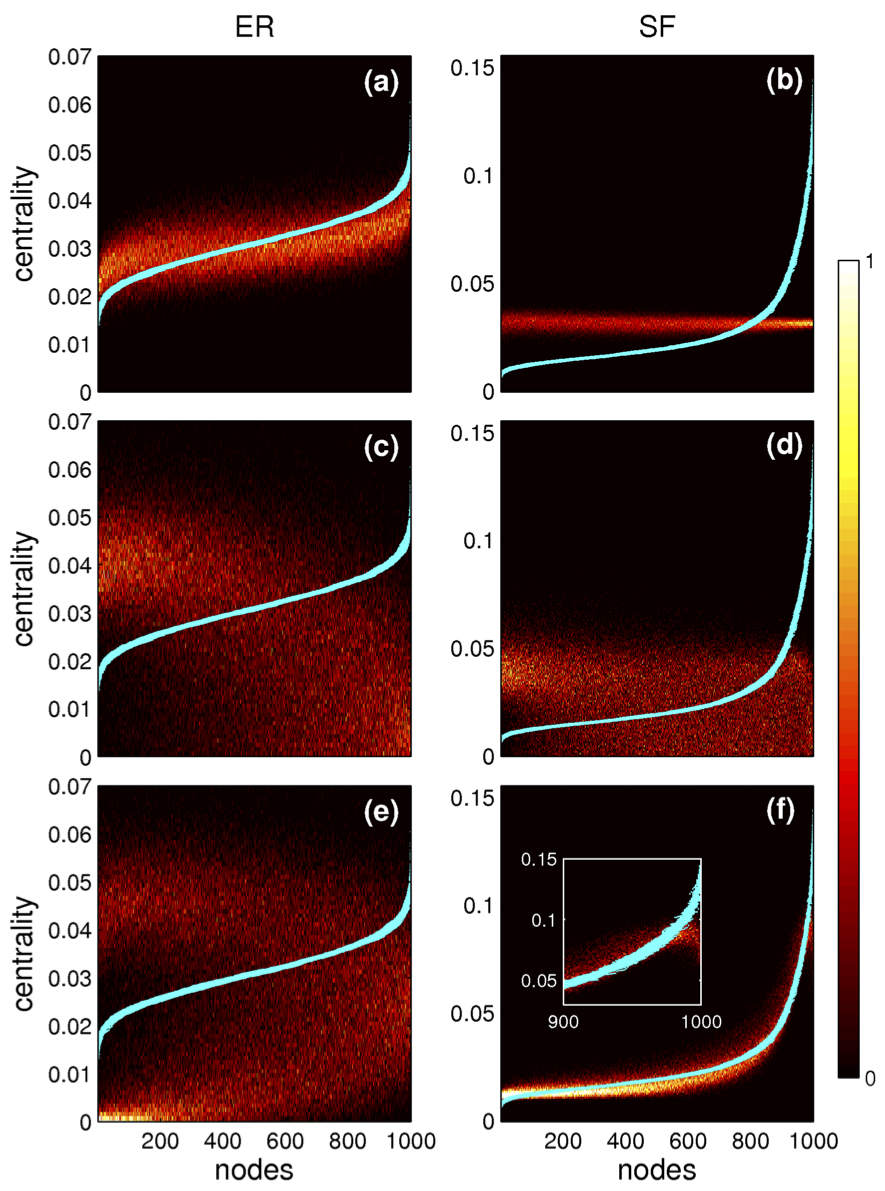}
\caption{(Color online). Synchronization centrality of explosive networks ($\Lambda_i^{\widetilde{C}}$, red clouds) versus that of non-explosive networks ($\Lambda_i^C$, cyan clouds) for ER  (left panels) and SF (right panels) topologies (see text for definitions). (a) and (b) account for the weighted method of Ref. \cite{Leyva13_w} with a uniform natural frequency distribution in $[-0.5, 0.5]$. (c) and (d) correspond to the weighted method of Ref. \cite{Zhang13} with the same uniform frequency distribution (see text for the methods' description). Panel (e) is the same as in (d) but natural frequencies are here chosen from the positive values of a Gaussian distribution, in a way that ES is no longer induced. Finally, (f) corresponds to a SF network where the degree-frequency correlation method from Ref. \cite{Gardenes11} is applied and the same uniform frequency distribution as for panels (a-d) is used. The inset in (f) is a zoom centered on the highest degree nodes. All data refer to averages over 100 realizations.}
\label{centrality}
\end{figure}

To show how the different studied procedures impact the synchronization centrality vector, we compare the original $\bf\Lambda^{C}$ obtained from Eq.~(\ref{C}) with the  $\bf\Lambda^{\widetilde{C}}$ resulting from the corresponding modified effective network $\widetilde C_{ij}=\Omega_{ij} (1-\frac{\Delta\omega_{ij}}{\Delta\omega_{max}})$.  Results for some of the different methods are condensed in Fig.~\ref{centrality} for ER (left panels) and SF (right panels) networks. In each panel, $\Lambda_i^{\widetilde{C}}$ (red dots) are plotted  together with $\Lambda_i^C$ (cyan dots) to show how the explosive method actually modified the  synchronization centrality vector and,
therefore, the dynamical role of each node. Nodes are sorted in ascending order of $\Lambda_i^C$. In all the cases where the structural and dynamical correlations introduced through $\Omega_{ij}$ successfully lead to ES  (Figs.~\ref{centrality}(a-d,f)), there is an increase (decrease) of $\Lambda_i^{\widetilde{C}}$ of those nodes whose $\Lambda_i^{C}$ was low (high), that is, the weighting method produces a flattening of $\Lambda_i^{\widetilde{C}}$.  In this way, the potential ability of the nodes to behave as seeds of local synchronization is frustrated until a certain coupling strength is reached. Only once the coupling strength is large enough, the rest of the network  fulfills the condition (\ref{condition}), and therefore a sudden transition to synchronization takes place.

More in detail,  Figs.~\ref{centrality}(a-b) correspond to the method described in Ref.~\cite{Leyva13_w}, where ES  is achieved choosing $\Omega_{ij} =A_{ij}|\omega_i-\omega_j|$ for ER networks (a) and $\Omega_{ij} =A_{ij}|\omega_i-\omega_j|l_{ij}/\sum_j l_{ij}$ for SF networks (b), being $l_{ij}$  the edge betweenness \cite{Boccaletti06}. Figures ~\ref{centrality}(c-d) show, instead,  the case $\Omega_{ij}=A_{ij}|\omega_i|/k_i$ proposed in Ref.~\cite{Zhang13} for uniform frequency distributions centred in zero. It is easy to see that the above increase-decrease compensation is fulfilled for both ER (c) and SF (d) networks. As expected, in SF networks the modification affects mainly the hubs, thwarting their dynamical influence as seeds, and frustrating the PTS. Finally, Fig.~\ref{centrality}(f) reports the case of ES induced in SF networks by imposing a frequency-degree correlation $\omega_i=k_i$ \cite{Gardenes11}. Here the effect is focused on the hubs (see the inset), whose synchronization centrality is now strongly decreased, while the imposed correlation does not produce a substantial difference between $\bf\Lambda^{\widetilde{C}}$ and $\bf\Lambda^{{C}}$ for the rest of the nodes.
There are however cases when, even if the structure and dynamics are correlated, ES does not occur.
For instance, Fig.~\ref{centrality}(e) reports the same case presented in Fig.~\ref{centrality}(c) but for positive definite frequencies. And indeed, for this particular frequency distribution, it is seen that the weighting method fails to flatten sufficiently the synchronization centrality (the red horseshoe cloud), with the consequence that ES fails to emerge as well.

An important application of our measure is the possibility of engineering an efficient strategy to produce ES in a generic network by {\it only}
acting upon a small fraction of its nodes, according to a given ranking defining their role as synchronization seeds. We here test four possible rankings:
\textit{i)} the \textit{synchronization ranking} based on $\Lambda^C_i$, \textit{ii)} the \textit{distance ranking}, that sorts the nodes according to the distance $\Delta\Lambda^C_i=\vert\Lambda^C_i- \Lambda^{\widetilde{C}}_i\vert$, \textit{iii)} the \textit{topology ranking}, based on $\Lambda^A_i$ and, finally, \textit{iv)} a random ranking, which is used for comparison. These specific rankings are actually suggested by the characteristic PTS occurring in both ER and SF networks, where the increase-decrease condition and the dominant role of hubs constitutes, respectively, the essential feature (see Fig.~\ref{centrality}).

\begin{figure}[t]
\centering
\includegraphics[width=0.49\textwidth]{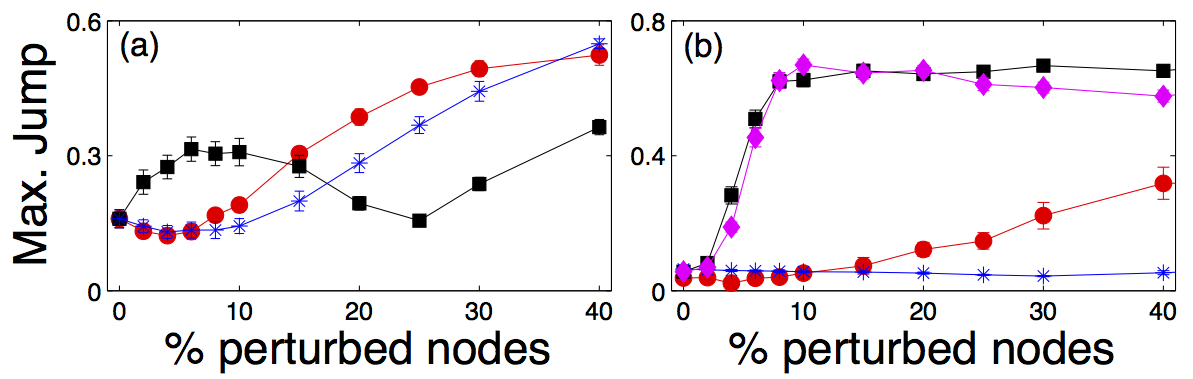}
\caption{(Color online). Maximum jump size in the synchronization curve $\sigma(r)$ vs. fraction of perturbed nodes chosen along several rankings: synchronization (back $\blacksquare$), distance (red $\bullet$), topology (magenta $\blacklozenge$) and random (blue $*$). (a) ER networks $\langle k\rangle=30$, weighting method from Ref.~\cite{Leyva13_w}. (b) SF networks $\langle k\rangle=6$ imposing $\omega_i=k_i$ for the selected nodes as in Ref. \cite{Gardenes11}. In all cases, data refer to averages over 50 realizations.}
\label{ranking}
\end{figure}

Figure \ref{ranking} reports the largest jump in the synchronization curve $r(\sigma)$ when several percentages of nodes (properly selected according to the ranking specified in the legend) are affected by the weighting methods of Refs. \cite{Leyva13_w,Gardenes11}.
For homogeneous ER networks (Fig.~\ref{ranking}(a)), we choose the method from Ref.~\cite{Leyva13_w, prescription}. The synchronization ranking (black squares) indicates that a significant explosive effect is obtained in the network already for just 6$\%$ of the nodes, whereas the distance ranking (red circles) requires up to 15$\%$ to get an equivalent jump, inducing a complete explosive transition only for percentages larger than 30$\%$. In comparison, using a random ranking (blue stars), it is necessary to manipulate at least the $40\%$ of the nodes.
For heterogeneous {SF} networks (Fig.~\ref{ranking}(b)), we apply instead the degree-frequency correlation method \cite{Gardenes11}, choosing a percentage of nodes and setting their natural frequencies equal to their degree. In this case, both the synchronization (black squares) and topological (magenta diamonds) rankings clearly outperforms the distance ranking (red circles) by only affecting the 10$\%$ of the nodes, while the distance ranking behaves slightly worst for large percentages. In comparison, the random ranking is not able to induce ES even for percentages above the $40\%$ \cite{Gardenes11}.
The differences between the two cases are mainly due to the different ways the seeds are spread in the network.
In the homogeneous case, the synchronization ranking obtains an optimum effect for small percentages since it focuses on the
seeds of synchronization. As soon as this percentage increases, the nodes with lowest synchronization centrality
are not captured by such a ranking, and the increase-decrease condition is not fulfilled.
As there are multiple randomly distributed seeds, the distance ranking is only slightly better than the random one,
as both satisfy the increase-decrease condition once the percentage is large enough.
In the heterogeneous case the seeds are just a few hubs, allowing to induce ES acting upon a very small fraction of the nodes
of the network, whereas the random targeting is definitely not the suitable choice.

In conclusion, we have introduced an effective network whose topological properties characterize quantitatively the PTS of networked oscillators, connecting the microscopic and the macroscopic behavior of the system during the process of
synchronization. With our approach we can reveal the inner mechanisms beneath ES,
which is shown to be rooted in a frustration of the PTS. This approach also provides a very simple yet effective way to predict whether or not a dynamical network synchronizes explosively for most of the various weighting procedures so far considered. Finally, synchronization centrality allows to induce such behavior locally, since we have the means to identify and isolate those seeds involved in the emergence of synchronization.

Authors acknowledge the computational resources and assistance provided by CRESCO, the supercomputacional center of ENEA in Portici, Italy.
Work partly supported by the Spanish Ministerio de Ciencia e Innovaci\'on  under projects FIS2012-38949-C03-01 and FIS2013-41057-P, and by the INCE Foundation under the project INCE2014-011.



\begin{thebibliography}{99}
\bibitem{prescription}
We choose the percentage  of nodes sorted by the corresponding ranking  and weight their links according to the rule $A_{ij}\rightarrow\Omega_{ij}/\overline{\Omega}_{ij}$, where $\Omega_{ij}=A_{ij}|\omega_i-\omega_j|$ and $\overline{\Omega}_{ij}$ is the mean of the non-zero elements of the weighted matrix. This normalization factor guarantees the presence of both reinforced and weakened links in the network, with the weighted adjacency matrix having values equal, larger, and lower than one.
\end{thebibliography}
\end{document}